# Seismic Fault Preserving Diffusion

1   **Olivier Lavialle**♠(1)(2), Sorin Pop(1), Christian Germain(1)(2), Marc Donias(1), Sebastien Guillon(3), Naamen Keskes(3),

2   Yannick Berthoumieu(1)

3
4   (1)   LASIS, Equipe Signal et Image UMR LAPS 5131 Av. Du Dr. Schweitzer BP 99, 33402 Talence Cedex, France
5         Tel : + 33 5 56 84 23 41 / Fax : + 33 5 57 35 07 79
6   (2)   ENITA de Bordeaux, BP 201, 33175 Gradignan Cedex, France
7         Tel : + 33 5 57 35 07 31 / Fax : + 33 5 57 35 07 79
8   (3)   SISMAGE, TOTAL, CSTJF Avenue Larribau, 64000 Pau, France
9         Tel : + 33 5 59 83 55 13 / Fax : +33 5 59 83 43 84
10
11
12

13  *Abstract*—This paper focuses on the denoising and enhancing of 3-D reflection seismic data. We propose

14  a pre-processing step based on a non linear diffusion filtering leading to a better detection of seismic faults.

15  The non linear diffusion approaches are based on the definition of a partial differential equation that allows

16  us to simplify the images without blurring relevant details or discontinuities. Computing the structure

17  tensor which provides information on the local orientation of the geological layers, we propose to drive the

18  diffusion along these layers using a new approach called SFPD (Seismic Fault Preserving Diffusion). In

19  SFPD, the eigenvalues of the tensor are fixed according to a confidence measure that takes into account the

20  regularity of the local seismic structure. Results on both synthesized and real 3-D blocks show the

21  efficiency of the proposed approach.

22  *Keywords*—3-D filtering, anisotropic diffusion, confidence measure, seismic data, structure tensor.

♠ corresponding author : ENITA de Bordeaux, BP 201, 33175 Gradignan Cedex, France  - Tel : + 33 5 57 35 07 31 / Fax : + 33 5 57 35 07 79  email : lavialle@enseirb.fr

## 1. Introduction

The acquisition and processing of reflection seismic data result in a 3D seismic block of acoustic impedance interfaces. The interpretation of these data represents a delicate task. Geological patterns are often difficult to recognize for the expert.

This interpretation of seismic blocks mainly consists in reflector picking (i.e. identifying and recording the position of specific reflection events) and fault plane locating. To be able to pick the reflectors wherever they are located throughout the seismic volume, the interpreter must be able to determine the vertical displacement across faults, and above all, he must discriminate whether a discontinuity is due to noise or artefacts or is evidence of a fault (Fig. 1).

As manual interpretation is both costly and subjective, some authors have investigated the use of image processing to develop automatic approaches (Admasu and Toennies, 2004; Randen et al, 2001; Sønneland et al, 2000). The resulting automatic tools are useful for structural interpretation of seismic data, but these tools failed in tracking horizons across faults especially if the level of noise is high.

One way to improve the efficiency of both manual and automatic interpretation is to increase the quality of the 3D seismic data by enhancing the structures to track as preserving the faults.

Among the different methods to achieve the denoising of 2D or 3D data, a large number of approaches using non-linear diffusion techniques have been proposed in the recent years (Weickert, 1997). These techniques are based on the use of Partial Differential Equations (PDE).

The simplest diffusion process is the linear and isotropic diffusion that is equivalent to a convolution with a Gaussian kernel.

The similarity between such a convolution and the heat equation was proved by Koenderink (1984):

$$\frac{\partial U}{\partial t} = c\Delta U = div(c\nabla U) \qquad (1)$$

In this PDE, $U$ represents the intensity function of the data; $c$ is a constant which, together with the scale of observation $t$, governs the amount of isotropic smoothing. Setting $c=1$, (1) is equivalent to convolving the image with a Gaussian kernel of width $\sqrt{2t}$. $div$ indicates the divergence operator.

Nevertheless, the application of this linear filter over an image produces undesirable results, such as edge and relevant details blurring.

To overcome these drawbacks Perona and Malik (1990) proposed the first non-linear filter by replacing the constant $c$ with a decreasing function of the gradient, such as:

$$g(|\nabla U|) = \frac{1}{1+(|\nabla U|/K)^2} \qquad (2)$$

where $K$ is a diffusion threshold. The diffusion process is isotropic for contrast values under the threshold $K$; gradient vector norms higher than $K$ are producing edge enhancing. Despite the quite convincing practical results, certain drawbacks remain unsolved: staircase effect (Whitaker and Pizer, 1993) or pinhole effect (Monteil and Beghdadi, 1999) are often associated with the Perona Malik process. In addition, in the strongly noised regions, the model may enhance the noise. Since the introduction of this first non-linear filter, related works attempted to improve it (Catte et al, 1992).

Weickert (1994; 1995) proposed two original models with tensor based diffusion functions. The purpose of a tensor based approach is to steer the smoothing process according to the directional information contained in the image structure. This anisotropic behaviour allows for adjusting the smoothing effects according to the direction.

The general model is written in PDE form, as:

$$\frac{\partial U}{\partial t} = div(D \nabla U) \tag{3}$$

with some initial and reflecting boundary conditions.

In the Edge Enhancing Diffusion (EED) model, the matrix $D$ depends continuously on the gradient of a Gaussian-smoothed version of the image ($\nabla U_\sigma$). The aim of this Gaussian regularization is to reduce the noise influence, having as result a robust descriptor of the image structure. For 2D application, the diffusion tensor $D$ is constructed by defining the eigenvectors ($\vec{v_1}$) and ($\vec{v_2}$) according to $\vec{v_1} \| \nabla U_\sigma$ and $\vec{v_2} \perp \nabla U_\sigma$ (Weickert, 1994). The corresponding eigenvalues $\lambda_1$, $\lambda_2$ were chosen as:

$$\begin{cases} \lambda_1 = \begin{cases} 1, & if \quad |\nabla U_\sigma| = 0 \\ 1 - \exp(\frac{-1}{|\nabla U_\sigma|^2}), & else \end{cases} \\ \lambda_2 = 1 \end{cases} \tag{4}$$

In this manner, EED driven processes are smoothing always along edges ($\lambda_2 = 1$) and, in the direction of the gradient, the diffusion is weighted by parameter $\lambda_1$ according to the contrast level in that direction.

Besides the EED model which enhances edges, Weickert proposed also a model for enhancing flow-like patterns: the Coherence Enhancing Diffusion - CED - (Weickert, 1999). The structure tensor introduced in this model is a powerful tool for analyzing coherence structures. This tensor $J_\rho$ is able to measure the gradient changes within the neighbourhood of any investigated point:

$$J_\rho(\nabla U_\sigma) = K_\rho * (\nabla U_\sigma \otimes \nabla U_\sigma) \tag{5}$$

Each component of the resulted matrix of the tensor product ($\otimes$) is convolving with a Gaussian kernel ($K_\rho$) where $\rho \gg \sigma$. The eigenvectors of $J_\rho$ represent the average orientation of the gradient vector ($\vec{v_1}$) and the structure orientation ($\vec{v_2}$), at scale $\rho$. The diffusion matrix $D$ (3) has the same eigenvectors as $J_\rho$, but its eigenvalues are chosen according to a coherence measure. This measure is proposed as the square difference between the eigenvalues of the structure tensor. The diffusion process acts mainly along the structure direction and becomes stronger as the coherence increases. In this manner, the model is even able to close interrupted lines.

Due to the characteristics of tensor $D$ (symmetry and positive eigenvalues), well posedness and scale-space properties were proved for both EED and CED models.

Based on these classical approaches, Terebes et al. (2002) proposed a new model, which takes advantage of both scalar and tensor driven diffusions. The mixed-diffusion combines the CED model with an original approach of the Perona Malik filter. The model aims at using the anisotropic diffusion in case of linear structures and a scalar diffusion otherwise. In order to avoid the development of false anisotropic structures and corner rounding (caused by the CED model), the scalar diffusion is applied to the regions with a noisy background and to junctions. The decision between types of diffusion is taken with respect to the global confidence proposed by Rao (1990). A strictly tensorial approach where the amount of diffusion was weighted by a sigmoid function depending on the Rao confidence is proposed in Terebes et al (2005).

Concerning the 3D applications, anisotropic diffusion has been frequently used in medical image processing. These works concern noise elimination (Gerig et al, 1992) but more often boundary detection and surface extraction (Krissian et al, 1995; Dosil and Pardo, 2003). Recently, specific PDE-based approaches were devoted to the seismic images filtering (Dargent et al, 2004a; Dargent et al, 2004b).

As we have seen, in most approaches an adaptive behaviour is obtained taking into account the local image structure and more particularly the local orientation. Concerning the characterization of the local structure, we have to mention the advanced works based on filter banks. These tools have proven efficient for orientation analysis (Granlund and Knutsson, 1995). In particular the first efficient approach was the steerable filters proposed by Freeman and Adelson (1991). Van Ginkel et al (1997) introduced an original deconvolution scheme leading to a better angular resolution of a Gaussian filter. Martens (1997) presented an application concerning the anisotropic noise reduction based on the sampled Hermite transform, efficient to represent 1-D structures in image. More recently, Gauthier et al (2005) proposed an application of a particular type of filter bank called "Complex Lapped Transform" for seismic data filtering. In this last case, due to the computational cost and the non-separability of the proposed transform, the application concerns only 2D slides of a 3D-Block. In addition, the authors conclude that the results have to be improved in term of fault preserving.

Furthermore, another non-PDE-based technique dedicated to the denoising of seismic structures was proposed by Bakker et al (1999). The authors combine edge preserving filtering with adaptive orientation filtering. The adaptive orientation filter consists in an elongated Gaussian filter steered by the eigenvectors of the structure tensor. Besides, a generalized Kuwahara filter, in which the window with higher confidence value is taken as a result, is proposed as edge preserving filter. This method leads to an enhancement of the faults when applied over the seismic images, but this enhancement is accompanied by a strong modification of the seismic data. We can note that one interest of this approach lies in its low computational cost.

In this paper we present a new approach based on the CED model, dedicated to 3-D seismic blocks processing. Seismic data are composed of strongly oriented patterns - stacks of almost parallel surfaces broken by faults. The aim of our method is to deliver a 3-D accurate image, from the fault detection point of view. So, our filtering consists in a data pre-processing method, which takes into consideration the enhancing of relevant discontinuities.

In section II we present the general 3-D CED model and some specific improvements for seismic data. A measure will be chosen to steer the diffusion along different coherence structures, such as plane-like or line-like structures. Relevant results, for both synthetic and real images, will be illustrated in section III. Finally, conclusions and further work will be presented.

**2.   Seismic data enhancing using 3-d anisotropic diffusion**

In this section, we present the extension of CED model in the 3-D case. Thanks to a confidence measure, we propose some improvements of this filter with respect to our type of data.

*A.  3-D CED model*

The 3-D model is a particular case of the general CED model (Weickert, 1995).

The structure tensor (5), becomes:

$$J_\rho(\nabla U_\sigma) = K_\rho * \begin{pmatrix} \left(\dfrac{\partial U_\sigma}{\partial x}\right)^2 & \dfrac{\partial U_\sigma}{\partial x}\dfrac{\partial U_\sigma}{\partial y} & \dfrac{\partial U_\sigma}{\partial x}\dfrac{\partial U_\sigma}{\partial z} \\ \dfrac{\partial U_\sigma}{\partial x}\dfrac{\partial U_\sigma}{\partial y} & \left(\dfrac{\partial U_\sigma}{\partial y}\right)^2 & \dfrac{\partial U_\sigma}{\partial y}\dfrac{\partial U_\sigma}{\partial z} \\ \dfrac{\partial U_\sigma}{\partial x}\dfrac{\partial U_\sigma}{\partial z} & \dfrac{\partial U_\sigma}{\partial y}\dfrac{\partial U_\sigma}{\partial z} & \left(\dfrac{\partial U_\sigma}{\partial z}\right)^2 \end{pmatrix} \qquad (6)$$

The smoothed version of intensity ($U_\sigma$) is obtained after a convolution with a 3-D Gaussian kernel:

$$K_\sigma(u) = \frac{1}{(2\pi\sigma^2)^{3/2}} \cdot \exp\left(-\frac{u^2}{2\sigma^2}\right) \qquad (7)$$

The *noise scale* ($\sigma$) establishes the minimum size of the objects preserved in the smoothed image. An average of the orientation, at *integration scale* $\rho$, is applied to deliver the orientation of the significant structures. Usually, the integration scale is chosen larger than the noise scale.

Due to the structure tensor properties (symmetric positive semi-definite), the eigenvalues are real and positive. These may be ordered as follows:

$$\mu_1 \geq \mu_2 \geq \mu_3 \qquad (8)$$

The corresponding eigenvectors ($\vec{v_1}, \vec{v_2}, \vec{v_3}$) form an orthogonal system. The largest eigenvalue carries the contrast variation in the dominant orientation of the averaged gradient vector ($\vec{v_1}$). The orientation corresponding to the lowest contrast difference is indicated by the third vector ($\vec{v_3}$).

Weickert introduces this knowledge of orientation in the general anisotropic model (3). Matrix $D$ has the same eigenvectors as the structure tensor. The orientation of the diffusion is driven by these eigenvectors and the intensity of the process by the eigenvalues of $D$. The author proposes the following system for choosing the eigenvalues of matrix $D$:

$$\begin{cases} \lambda_1 = \lambda_2 = \alpha \\ \lambda_3 = \begin{cases} \alpha & \text{if } k = 0 \\ \alpha + (1-\alpha)\exp\left(\dfrac{-C}{k}\right) & \text{otherwise} \end{cases} \end{cases} \qquad (9)$$

The parameter $\alpha$ represents the amount of diffusivity in the orientations of the highest fluctuation contrast. In order to hamper the diffusion in these orientations, the parameter $\alpha$ is chosen nearly 0. For theoretical reasons this parameter must be positive.

The measure of coherence $k$ is defined as:

$$k = (\mu_1 - \mu_2)^2 + (\mu_1 - \mu_3)^2 + (\mu_2 - \mu_3)^2 \qquad (10)$$

The threshold parameter $C$ is usually chosen equal to 1. In the coherent structures ($k \gg C$), the diffusion processes essentially along $\vec{v}_3$ ($\lambda_3 \approx 1$). On the other hand, if the structure becomes isotropic ($k \to 0$), the amount of diffusivity in all three orientations is no more than $\alpha$.

This coherence measure $k$ depends on the gradient energy. For this reason, the amount of diffusivity ($\lambda_3$) in the third vector orientation always tends to 1. In conclusion, this system will smooth only in one orientation of space, which is not adapted to seismic data.

In order to deal with plane-like structures like seismic horizons, the first idea consists in forcing the diffusion process along both the second and third eigenvectors. We can easily obtain such a result by choosing a set of eigenvalues different from (9):

$$\begin{cases} \lambda_1 = \alpha \\ \lambda_2 = \lambda_3 = \begin{cases} \alpha & \text{if } k = 0 \\ \alpha + (1-\alpha)\exp\left(\dfrac{-C}{k}\right) & \text{otherwise} \end{cases} \end{cases} \qquad (11)$$

In the result section, the original weickert's approach will be denoted CED-1D and the approach based on this new set of eigenvalues will be denoted CED-2D as it allows filtering of 2D structures.

Using the CED-2D approach leads to a diffusion process steered along the 2D horizons even in the presence of faults. As a consequence, this method presents the drawback of smoothing the signal across faults leading to a loss of relevant seismic information.

Considering the behaviour of the CED-1D and CED-2D methods, we will propose a new approach which consists in choosing an appropriate set of eigenvalues to both enhancing the structures to track and preserving the faults as relevant details.

*B. Confidence measures for seismic data*

Among important features of seismic 3-D data, faults represent an interesting point for our treatment. A simplified view describes the seismic data like stacks of almost parallel planes (horizons) broken by faults. We may interpret these strongly oriented data as linear structures. Van Kempen et al (1999) define the notion of dimensionality of structures. In 3-D case, beside the isotrope structures corresponding to three shift invariant orientations, two types of linear structures are possible:

- plane-like linear structure – shift invariant along two orientations,
- line-like linear structure – shift invariant along one orientation.

Analysis of the linear structure may be issued by the computation of the structure tensor. Thus, the vectors of the structure tensor point out the principal axes of orientation and the number of the zero eigenvalues indicates the number of the shift invariant orientations.

In the seismic case, the horizons can be viewed as plane-like structures. A horizon is characterized by a large eigenvalue and two others close to zero. A fault is characterized by two large eigenvalues and the other close to zero. This property is due to the fact that the orientation of the average gradient around the fault is a mixture of two distinct orientations corresponding to the neighbourhood regions. Thus, we can model the fault as a line-like structure, although, from a seismic point of view, it is rather a plane than a line.

In 2001, Bakker et al, following the works of Bigun et al (1991), proposed two measures to estimate the semblance of seismic data with this type of linear structures:

$$C_{plane} = \frac{\mu_1 - \mu_2}{\mu_1 + \mu_2} \qquad C_{line} = \frac{\mu_2 - \mu_3}{\mu_2 + \mu_3} \qquad (11)$$

These measures are combined to obtain a fault confidence:

$$C_{fault} = C_{line}(1 - C_{plane}) \tag{12}$$

We can note that the author proposes to introduce this measure as a confidence value to select the optimal mask in an approach combining orientation adaptive filtering and edge preserving filtering. The introduction of such a priori measures leads to a technique which strongly enhances the detected faults. The more serious drawback of this approach is that it leads to a too important data transformation.

*C. Seismic Fault Preserving Diffusion*

We propose a new approach of the general CED model, more appropriate for seismic data. Keeping the general equation of the anisotropic diffusion (3) we introduce an adaptive system to fix the $D$ matrix eigenvalues.

We intended to create a system adapted to local context, which acts in specific ways for different regions. We chose the confidence measure $C_{fault}$ from the various set of measures dedicated to this purpose (Rao, 1990; Berthoumieu, 2006). The reason why we selected this type of measure is its closed link to the nature of our seismic data. We propose the following system:

$$\begin{aligned} \lambda_1 &= \alpha \\ \lambda_2 &= \lambda_3 - (\lambda_3 - \lambda_1) h_\tau(C_{fault}) \\ \lambda_3 &= \begin{cases} \alpha & \text{if } k = 0 \\ \alpha + (1-\alpha)\exp\left(\dfrac{-C}{k}\right) & \text{else} \end{cases} \end{aligned} \tag{13}$$

where $h_\tau(s)$ is described in (Terebes et al., 2005):

$$h_\tau(s) = \frac{\tanh[\gamma(s-\tau)]+1}{\tanh[\gamma(1-\tau)]+1} \tag{14}$$

215   The eigenvalue $\lambda_2$ depends continuously on the confidence measure ($C_{fault}$) and takes values

216   between $\lambda_1$ and $\lambda_3$. In the neighbourhood of a fault, $\lambda_2$ tends to $\lambda_1$ whereas it tends to $\lambda_3$ when $C_{fault} \to 0$.

217   Through the value of two parameters the threshold $\tau$ and the slope $\gamma$, the sigmoid function $h_\tau(s)$ allows a

218   better control of transition between two homogeneous regions.

219   Within presumptive fault zones ($C_{fault} \to 1$), the process will only smooth along the smallest variation of

220   contrast ($\vec{v}_3$). In this case the amount of diffusivity in the first and in the second orientation given by the

221   $\lambda_2$ and $\lambda_1$ values is equal to $\alpha$ chosen near to 0.

222   The regions where $C_{fault} \to 0$ are rather characterized by plane-like structures ($C_{plane} \to 1$). In these

223   regions the process will diffuse in the plane defined by the vectors $\vec{v}_2$ and $\vec{v}_3$. This plane is orthogonal to

224   the average gradient. For this type of horizons the coherence measure $k$ is high ($\mu_1 >> \mu_2 \approx \mu_3$) and forces

225   the $\lambda_2$ and $\lambda_3$ values to reach 1.

226   **3. Results**

227   This section illustrates the efficiency of our approach on both synthetic and real seismic blocks. The

228   noise reduction and the faults preserving are evaluated. Our filter is compared with both the CED-1D and

229   CED-2D models.

230   *3.1. 3D-synthesized blocks*

231   Since it is much easier to judge the efficiency of the algorithms on a synthetic image, we propose to use a

232   3-D synthetic block composed by a stack of layers with a sinusoidal profile and broken by two crossed

233   faults. Figure 2 shows a front section of the original block.

The data are corrupted with additive Gaussian white noise. Figure 3 shows the noisy blocks for signal-to-noise-ratio (SNR) of 1 dB, 3 dB and 5 dB. Each noisy block is filtered with our method and the CED methods. Parameters common to the various algorithms take on the same values ($dt=0.05, \sigma = 0.4, \rho = 1.2, \alpha = 10^{-4}$, *120 iterations*). In addition, parameters specific to SFPD are set to $\tau = 0.1$ and $\gamma = 10$. We show in Figure 4 the results obtained using an explicit numerical scheme.

Figure 5 shows the top views (i.e. time slices) corresponding to the SFPD and both CED results for the 3 dB noisy block.

The efficiency of our method was evaluated by the means of root-mean-square-error (RMSE) which allows quantifying the similarity between each diffused block and the original synthetic block:

$$RMSE = \sqrt{\frac{\sum_{x,y,z}(U(x,y,z) - U_0(x,y,z))^2}{n}} \qquad (15)$$

where $U_0$ denotes the value of the voxel with coordinates (x,y,z) in the original non-noisy block (Fig. 2) and $U$ the value of the same voxel in the processed image. $n$ denotes the total number of voxels.

Firstly, the original block was segmented in two regions: faults and non-faults. This segmentation was achieved using a simple thresholding on the $C_{fault}$ value. Then, for each processed block, the RMSE has been computed in these two different zones in order to illustrate the behaviour of the methods in particular in the fault regions. The resulting RMSE values are provided in Table 1.

Considering the quality of the denoising, our approach performs well when compared with the CED models, in terms of both visual quality and global RMSE. In particular, false anisotropic structures appear in the block processed with the CED 1D model (Fig. 5c) while our approach does not create this type of structure (Fig. 5d). This is also reflected in the RMSE values corresponding to the non-fault region.

Like our model, CED 1D preserve the faults producing comparable RMSE values in fault region, which is not the case of CED 2D model. On the other hand CED 2D model provides a good quality in the non-fault zones (Fig. 4c, 4f, 4i).

Finally, considering both the noise reduction and the fault preserving, we can conclude that the proposed SFPD model takes advantage of the 1D and 2D Weickert's models.

*3.2. Real 3D-reflection seismic data*

Figure 6 compare results generated on a real seismic block (Fig. 1) by CED 1D, CED 2D and SFPD respectively. These results illustrate that SFPD is better adapted to remove the noise while preserving the fault.

**4. Conclusions**

We have proposed a new approach of tensorial diffusion which takes into account the characteristics of seismic data. More precisely, we make sure that our denoising approach preserves the faults. For this purpose we use a measure of fault confidence in a tensor driven diffusion process adapted to the local context. This measure allows us to diffuse only in one orientation in a fault neighbourhood and to perform a diffusion process guided by two orientations along the layers otherwise. This approach also exempts from the creation of false anisotropic structures, artefacts typically observable in images processed with the classical tensorial models. Our method can be used as a preprocessing for automatic or manual interpretation of 3D reflection seismic data.

Future works will focus on improving our model by adding more seismic-data-specific properties.

| Original SNR values (dB) | Methods | RMSE | | |
|---|---|---|---|---|
| | | Fault regions | non-Fault regions | Whole block |
| 1.0 | SFPD | 14.548 | 7.560 | 8.569 |
| | CED 1D | 16.629 | 14.370 | 14.628 |
| | CED 2D | 18.648 | 7.412 | 9.564 |
| 3.0 | SFPD | 11.523 | 3.893 | 5.002 |
| | CED 1D | 11.681 | 7.837 | 8.247 |
| | CED 2D | 18.113 | 5.205 | 8.037 |
| 5.0 | SFPD | 10.930 | 2.835 | 4.067 |
| | CED 1D | 10.330 | 5.582 | 6.109 |
| | CED 2D | 18.058 | 4.622 | 7.691 |

Table 1. RMSE values for the diffusion of noisy synthesized 3D-blocks in both fault and non-fault regions.

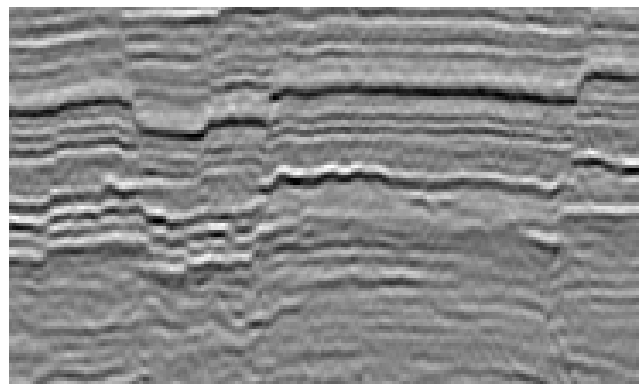

Figure 1: A section of 3D seismic data

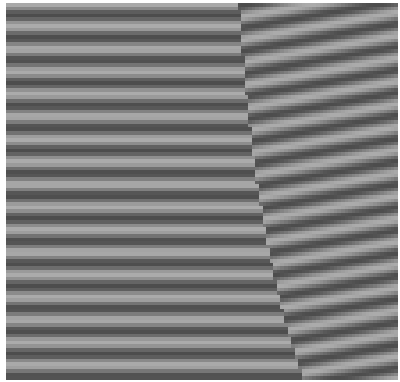

Figure 2: Front section of a synthesized block

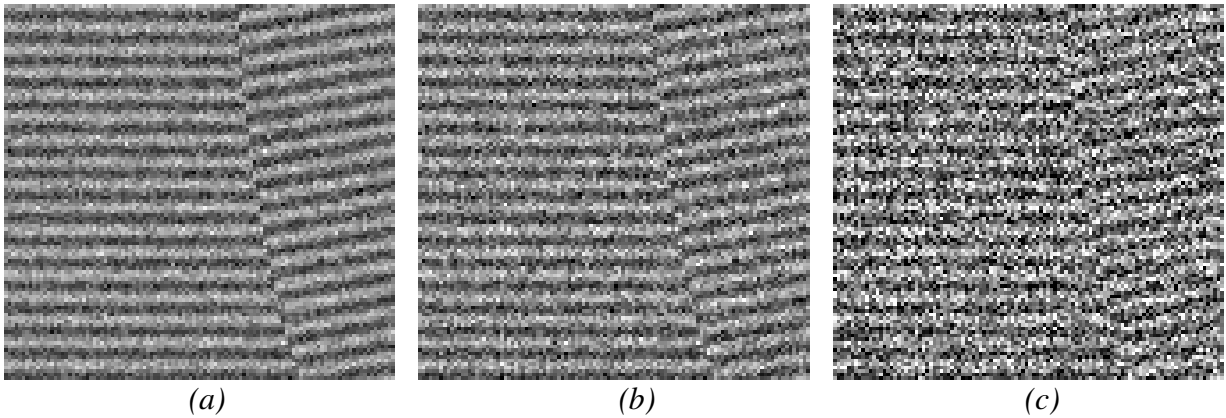

*(a)* *(b)* *(c)*

Figure 3: Front section of noisy synthesized 3D-blocks. SNR= (a) 5dB (b) 3dB (c) 1dB

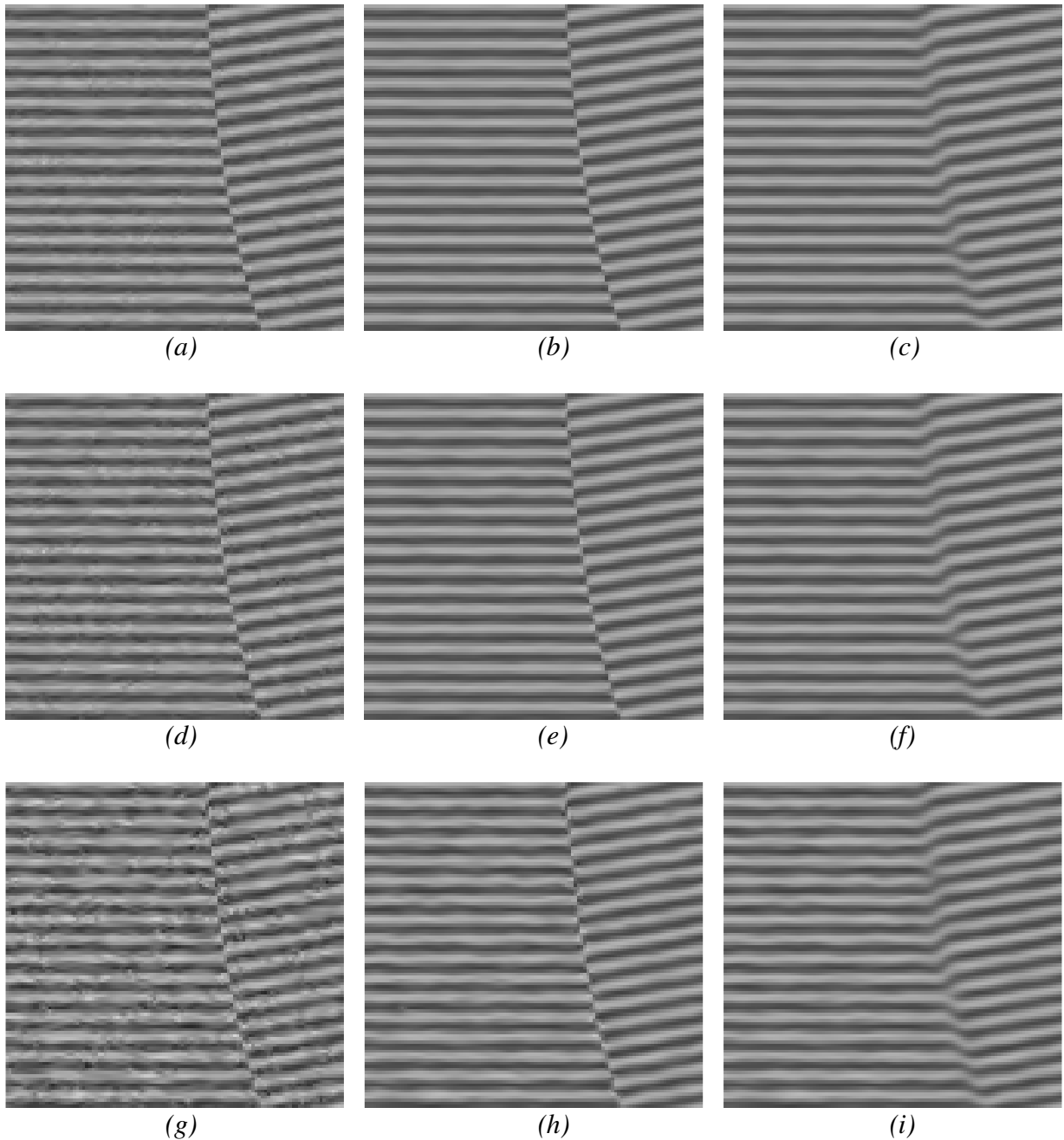

Figure 4: Diffusion Results for synthesized block. First row: diffusion of the noisy block with SNR=5dB (a) CED 1D, (b) SFPD, (c) CED 2D ; second row: diffusion of the noisy block with SNR=3dB: (d) CED 1D, (e) SFPD, (f) CED 2D ; third row: diffusion of the noisy block with SNR=1dB: (g) CED 1D, (h) SFPD, (i) CED 2D.

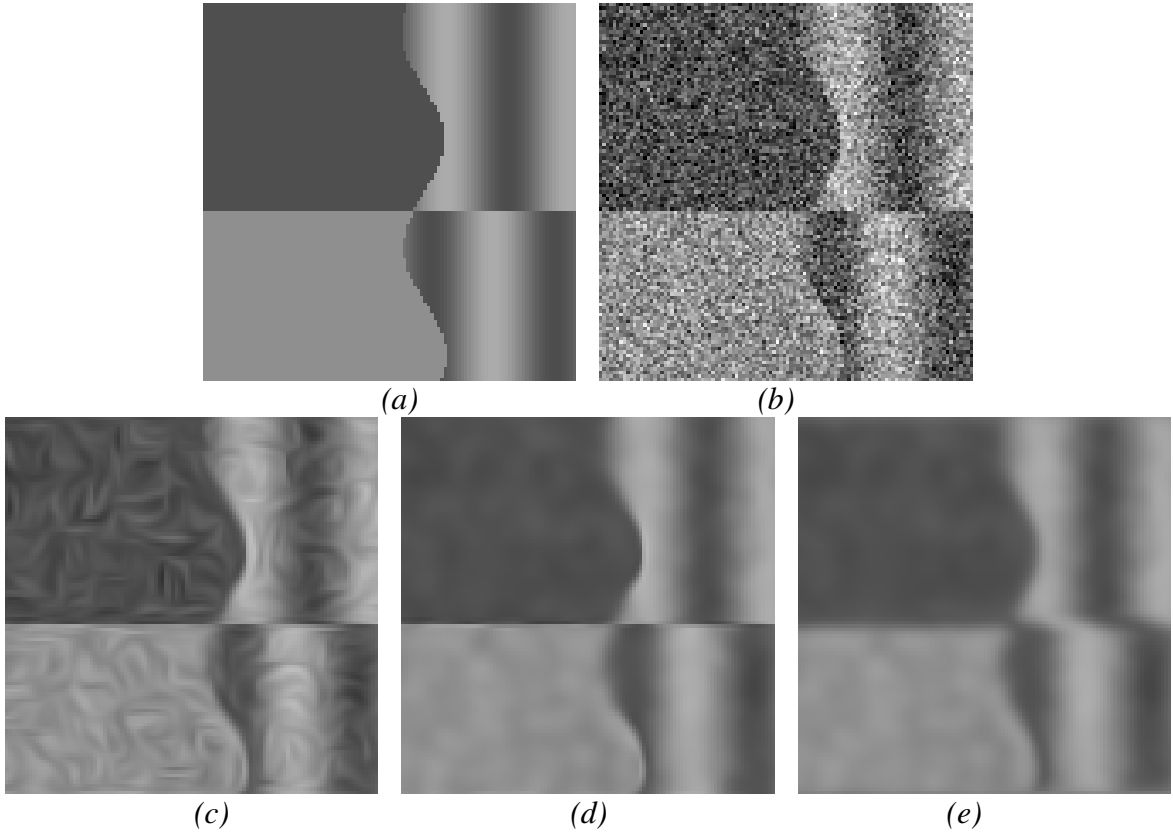

Figure 5: Top view of diffused blocks. (a) Original (b) noisy-SNR=3dB (c) CED 1D-diffusion (d) SFPD diffusion (e) CED 2D diffusion

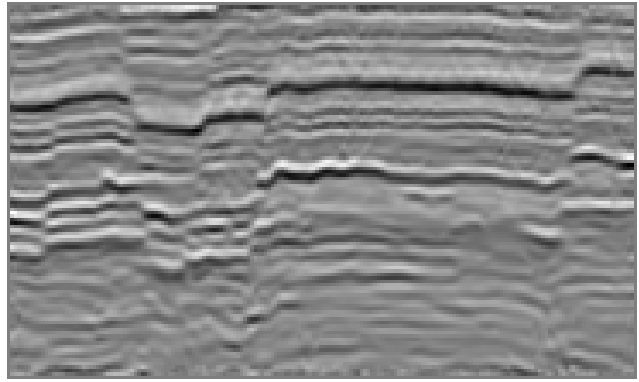
(a)

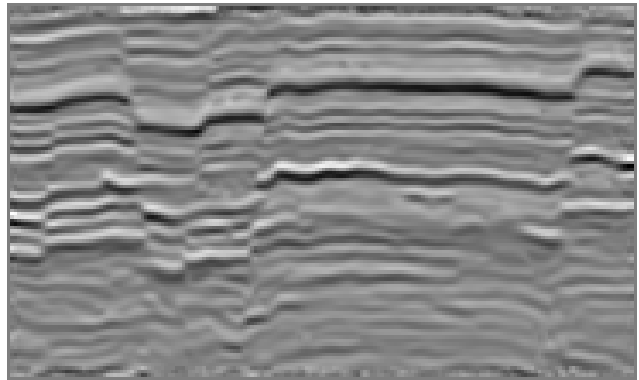
(b)

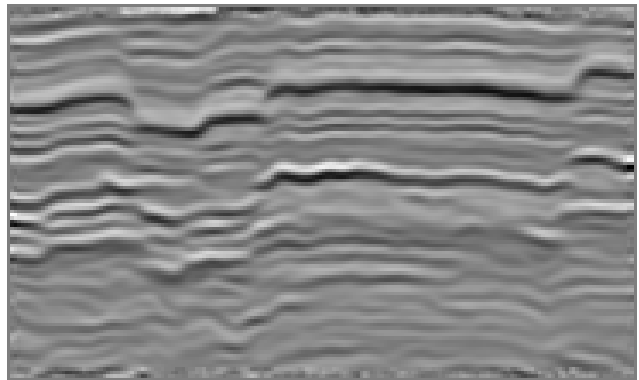
(c)

Fig.6 Diffusion Results for the real 3D seismic block. (a) CED 1D diffusion (b) SFPD-diffusion

(c) CED 2D-diffusion